\documentclass[letter]{aa} 

\usepackage{graphicx}
\usepackage{txfonts}
\usepackage{color,xspace}
\newcommand{\obj}{AGESVC1 282\xspace}
\newcommand{\HI}{H\textsc{i}}
\newcommand{\Msolar}{M$_\sun$}
\newcommand{\kms}{km\,s$^{-1}$}

\begin{document} 

    \title{Deep optical imaging of the dark galaxy candidate \obj }

   \author{Michal B\'ilek\inst{1}
          \and Oliver M\"uller\inst{1}
          \and Ana Vudragovi\'c\inst{2}
          \and Rhys Taylor\inst{3}}

   \institute{Universit\'e de Strasbourg, CNRS, Observatoire astronomique de Strasbourg (ObAS), UMR 7550, 67000 Strasbourg, France\\
 \email{bilek@astro.unistra.fr}
         \and
             Astronomical Observatory, Volgina 7, 11060 Belgrade, Serbia\\
        \and
            Astronomical Institute of the Czech Academy of Sciences, Bo{\v c}n\'i II 1401/1a, 141 00 Praha 4, Czech Republic\\
             }

   \date{Received ...; accepted ...}

 
  \abstract{The blind \HI{} survey Arecibo Galaxy Environment Survey (AGES) detected several unresolved  sources in the Virgo cluster, which do not have optical counterparts in the Sloan Digital Sky Survey. The origin of these dark clouds is unknown. They might be crucial objects since they could be the so-called dark galaxies, that is, the dark matter halos without stellar content that are expected from cosmological simulations. 
   In order to reveal the nature of the dark clouds, we took a deep optical image of one them, \obj, with the newly-commissioned 1.4\,m Milankovi\'c Telescope. After observing it for 10.4\,h  in the  $L$-filter, the image reached {a} surface-brightness limit of about {29.1}\,mag\,arcsec$^{-2}$ in $V$. No optical counterpart was detected.
   We {placed} an upper limit on the $V$-band luminosity of the object of $1.1\times10^7\,L_\sun$, {giving a} stellar mass below {$1.4\times10^7\,M_\sun$ {and} a \HI{}-to-stellar mass ratio above 3.1}. {By inspecting archival \HI{} observations of the surrounding region, we found that none of the standard explanations for optically dark \HI{} clouds fits the available constraints on this object.}}
  
               
   \maketitle
%

\section{Introduction}
Recently, there has been a rise of interest in the low surface brightness (LSB) universe. While the existence of LSB galaxies has been long established (e.g., \citealt{s84}, \citealt{malin1}); in the last {few} years, it has become possible to detect them in large numbers (e.g., \citealt{vandok,koda,2017A&A...608A.142V,2017A&A...597A...7M,2017MNRAS.468..703R,2018ApJ...857..104G,2019MNRAS.488.2143P,2020MNRAS.491.1901H}). Given the difficulties in detecting these faint objects, it is natural to ask {whether} there might be a significant population of even fainter galaxies with perhaps no stellar content at all \citep{2020MNRAS.491.3496C}. {The} existence of dark matter halos without stellar content is the standard way to explain the missing satellites problem of cosmological dark-matter-only simulations \citep{2007ApJ...670..313S,2016MNRAS.457.1931S,2018MNRAS.478..548S}.

One proposed way to detect such ``dark galaxies'' is through \HI{} surveys. Indeed, there appears to be a gas density threshold for star formation (see \citealt{dgmodels} for a detailed discussion). While numerous dark clouds, that is to say \HI{} sources without optical counterparts, have been detected, establishing their nature is often difficult: \HI{} removed through tidal encounters or ram pressure stripping might {have a similar morphology} and kinematics as dark galaxies (\citealt{duc}). However, there is a set of objects in the Virgo cluster that does not appear to {be} explicable by tidal encounters (\citealt{t17}) or through interactions with the intracluster medium (\citealt{t18}). Their velocity widths are consistent with a significant dark matter content, and simulations have shown that such dark galaxies would be able to withstand a disruption by other cluster members (\citealt{t16}).

These objects were first reported in \cite{t12} and \cite{t13} as part of the Arecibo Galaxy Environment Survey (AGES). Their radial velocities {(1000-2000\,\kms)}  and distribution on the sky suggest that they are members of the Virgo cluster {(in the case of \obj{} in the B cloud, at a distance of 23 Mpc, \citealt{gv99})}. Their \HI{} masses are $\sim$1-3$\times$10$^{7}$\Msolar{}, which is consistent with other tidal debris; however, their {combination of} isolation ($>$\,100 kpc from the nearest galaxy) and high line widths ($\sim$150\,\kms) make them more unusual. {Tidal debris is known to be found at large distances from its parent galaxy (e.g., \citealt{hess}, \citealt{leisman}, \citealt{serra}), but with line widths typically $<$\,50\,\kms{}. In contrast, high line width features are known to exist within large \HI{} streams (e.g., \citealt{koop}, \citealt{kent}), but they are rarely found without a clear association to a likely parent galaxy. In this particular respect, the AGES Virgo clouds are highly unusual.} In addition, their low mass and high velocity widths make them strongly deviant from the baryonic Tully-Fisher relation \citep[BTFR,][]{2005ApJ...632..859M} if at the distance of the Virgo cluster. In order to reconcile them with the usual BTFR, they would need distances of $\sim$2\,Mpc, but there are no other known foreground galaxies in the direction of the Virgo cluster. Analogous objects have not been detected in other galaxy clusters.

{In this paper, we describe extremely} deep optical imaging {to try} to detect the optical counterpart of one of these dark clouds, \obj. This can potentially bring several possible benefits (see Sect.~\ref{sect:disc} and also \citealt{VIRGOHI21}): We might discover that the object has a faint galaxy as an optical counterpart (as in \citealt{imp} or \citealt{mihos}), that the gas has been displaced from {a} faint parent galaxy that is much closer to the cloud than to the known brighter objects (as in the case of the \HI{} cloud near VCC\,1249, \citealt{vcc1249}), or, alternatively, we might find stellar tidal debris that would {indicate a} tidal origin of the clouds. No optical counterpart was actually detected. The luminosity of the object must be below  $1.1\times10^7\,L_\sun$.  In this paper, we assumed  H$_{0}$ = 71\,\kms\,Mpc$^{-1}$ for consistency with earlier AGES papers, {and we assume a distance to the object of 23 Mpc}.

\section{Observations and data reduction}
The \HI{} observations and data reduction of AGES are extensively described in \cite{auld}, \cite{t12}, and \cite{t13}. In brief, AGES is a drift scan survey with a sensitivity of~0.7\,mJy, a spatial resolution of 3.5\arcmin , and a spectral resolution of 10\,\kms {after Hanning smoothing}. This gives an approximate mass sensitivity of 8$\times$10$^{6}$\Msolar{} at the Virgo distance with a physical resolution of 23\,kpc in the B cloud of the cluster. {Objects in AGES were detected} through a combination of visual and automated source extraction procedures, and they were confirmed with follow-up observations using Arecibo's L-wide receiver. Eight optically dark clouds were discovered in Virgo by AGES. We selected \obj as our target (coordinates RA = 12h 25m 24.10s, dec =   +08\degr 16\arcmin 54\farcs0)  because {it has one of the highest \HI{} signal-to-noise ratios (11.4) and velocity widths} (W20 = 164 \kms{}), though these features are similar to those of the other clouds. The width of the Arecibo beam puts an upper limit on the \HI{} radius of \obj of 105\arcsec. The lower limit on the \HI{} radius is harder to establish, but it is likely higher than about 1\,kpc (10\arcsec); otherwise, the column density would exceed that of bright, star-forming galaxies  \citep{t16}. {Nevertheless this calculation is necessarily simplistic since predicting the optical brightness of the clouds is highly nontrivial (e.g., \citealt{bacch})}. The \HI{} mass of \obj was estimated as $10^{7.64}\,M_\sun$ (\citealt{t12}).

The target was observed during the two nights of 17 and 18 April 2020 with the 1.4\,meter Milankovi\' c telescope equipped with {an} Andor IKONL CCD camera from the Astronomical Station Vidojevica (Serbia). Several images were taken with a 90\degr \ camera rotation. {Sky} flat field images were taken for both regular and rotated camera positions. A total of 125 images were taken in the luminance $L$-filter with a 300\,s exposure each. 
The integrated exposure time was 10.4\,hrs.
The nights were clear with a $\sim1$\,hr period of thin cirrus. {We measured  an average image quality of 1\farcs4}.  Utilizing the focal reducer delivers a square field-of-view with sides of 13\farcm3 { with a pixel scale of 0.39 pixels per arcsec.} {We applied a large dithering pattern randomized within a maximum offset of 4.5\arcmin, which is crucial when digging deep into the low-surface brightness regime at the 28-29\,mag arcsec$^{-2}$ level \citep{2009PASP..121.1267S,2019arXiv190909456M} and beyond \citep{2016ApJ...823..123T}. Such a strategy provides a strong handle on the background subtraction, as well as on the removal of systematics arising from the instruments and observations, that is, from reflections or flat-fielding imperfections.} 

{The calibration of the images follows the standard procedure as described in \citet{2019A&A...632L..13M} and includes bias, dark, flat, and illumination corrections with a careful customized background subtraction. The python pipeline implementing this calibration can be found online\footnote{{gitlab.com/VoltarCH/milankovic-telescope-reduction-pipeline}}. To create our final scientific image, we stacked all calibrated science frames with SWarp \citep{2010ascl.soft10068B}, where we used a median for stack.
The final, fully calibrated, and dithering image reached a field-of-view of 21\farcm9 $\times$ 22\farcm7.  
The astrometry in the science frames was solved locally with the API of astrometry.net \citep{2010AJ....139.1782L}.} %

 \begin{figure}
     \centering
     \includegraphics[width=0.99\linewidth]{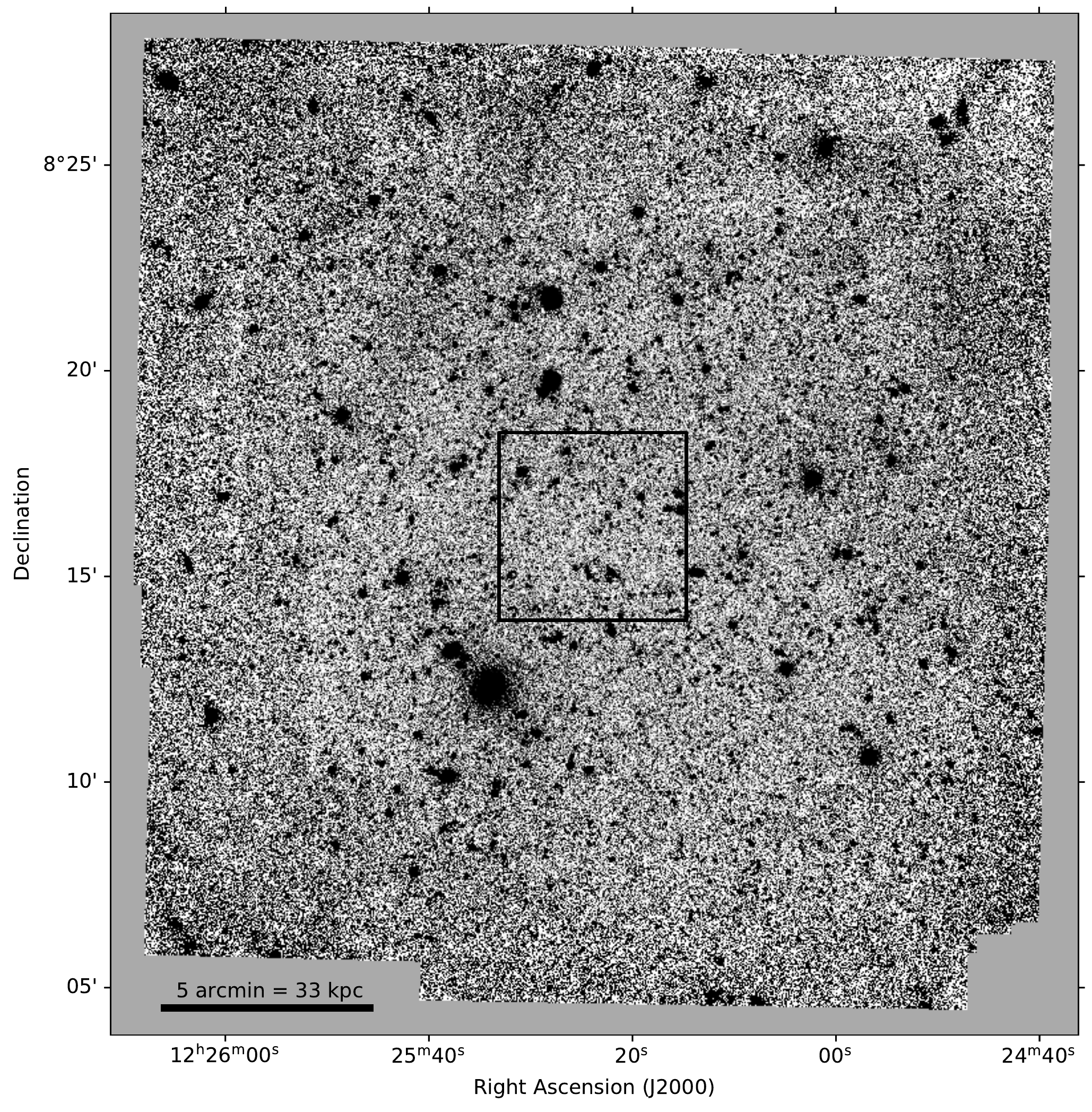}
     \caption{Fully stacked and calibrated image.  The image is centered on the dark cloud \obj. The field-of-view is 21\farcm9 $\times$ 22\farcm7. {The box indicates the stamps presented in Figure \ref{fig:artificial}. The black line shows the scale at an assumed distance of 23\,Mpc.}}
     \label{fig:artificial}
 \end{figure}

The zero point of this stacked image was estimated using the Sloan Digital Sky Survey (SDSS) DR12 star catalog \citep{2015ApJS..219...12A}. We used Source Extractor to measure isophotal magnitudes of the stars in the $L$-band.
{Since this catalog does not list stellar magnitudes in the $L$ filter, but instead the  SDSS $g$- and $r$-band magnitudes}, we follow \citet{2016A&A...588A..89J}, who applied a linear transformation between the $L$- and the $r$-band, thus taking into account a minor dependence on the $g-r$ color (see their Figure 1). 
{In cross-matching our instrumental magnitudes  to the SDSS DR12 catalog, we  found 350 stars in common that are brighter {than} 21 magnitude in the $g$-band. After the exclusion of stars that are too bright (saturated), we ended up with 336 stars. Furthermore, we decided to transform our magnitudes {from  the $g$- and $r$-bands to the $V$-band using the} transformations from \citet{2005AJ....130..873J}. This procedure yielded an intercept of 31.6 $\pm$ 0.1 mag in the $V$-band and a slope of 1.03 $\pm$ 0.01 in the AB system}.  From the standard deviation in 30 boxes with sizes of $10\times10$ arcsec$^2$, we estimated a $1\sigma$ surface brightness limit of $29.05\pm0.04$ mag arcsec$^{-2}$ in the $V$-band. The error of this limit was estimated by the standard deviation of the measured surface brightnesses of the different boxes.

 \begin{figure*}
     \centering
     \includegraphics[width=0.99\linewidth]{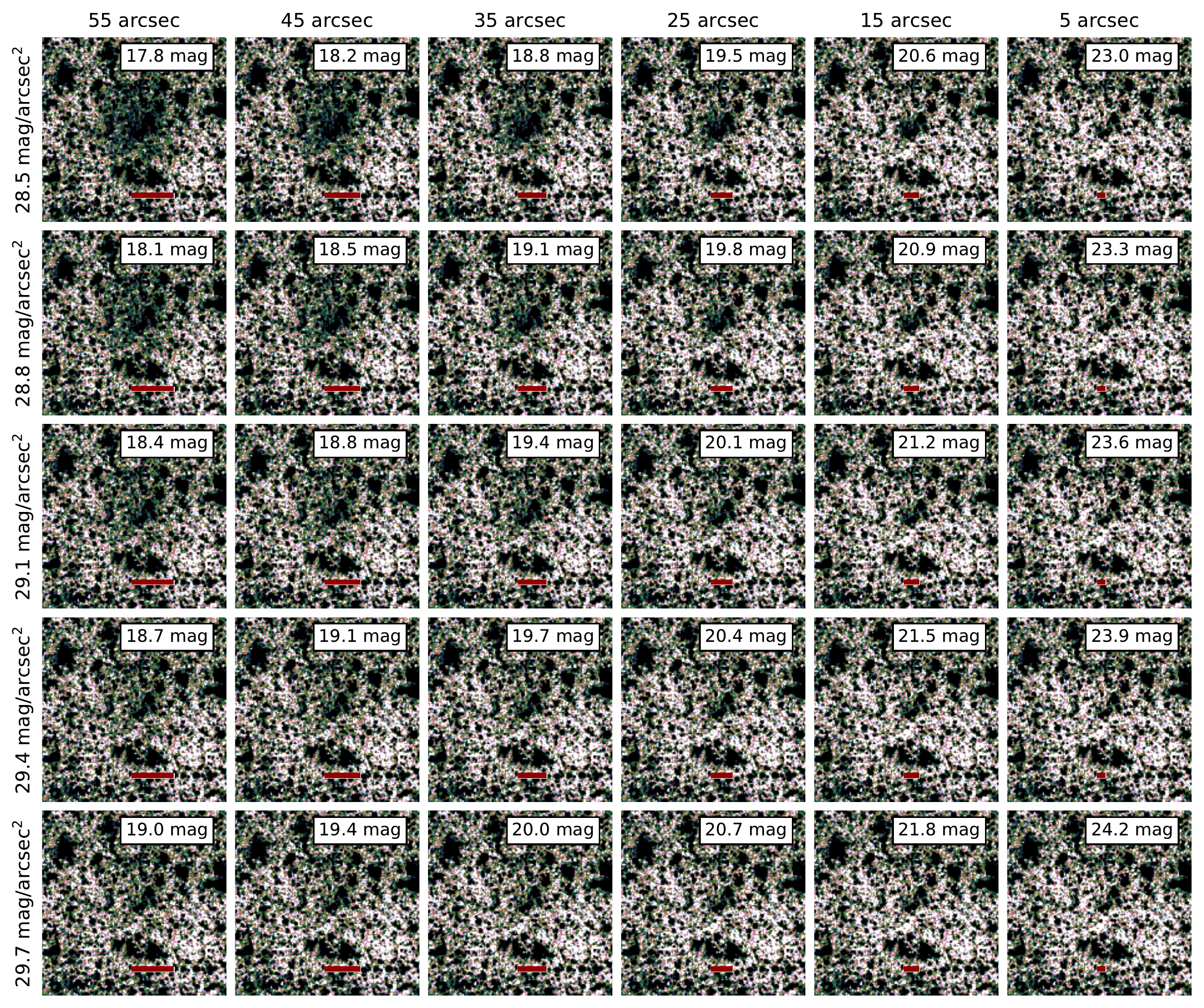}
     \caption{Injected artificial dwarf galaxies. 
     The effective radius and the effective surface brightness are varied in the horizontal direction and the vertical direction, respectively. The red line indicates the effective radius for each artificial galaxy. The number in the box represents the {total} apparent magnitude of the artificial galaxy. The images are  smoothed with {a Gaussian kernel with $\sigma = 2\,$px to enhance the LSB features. The brightness of the stamps is represented with a cube helix color scheme \citep{2011BASI...39..289G}.}}
     \label{fig:field}
 \end{figure*}

{To test the detectability of a LSB  galaxy  at the position of \obj, we injected a range of artificial dwarf galaxies into the image. For this purpose, we modeled them with a 2D S\'ersic profile with different effective radii and surface brightnesses. We used a S\'ersic index of $n=0.6$, which is typical for faint dwarf galaxies \citep[e.g., ][]{2017A&A...597A...7M,2017A&A...608A.142V}. Zero ellipticity was assumed. {We used as {maximum} effective radius of 6\,kpc, that is, the size of the most extended  low-luminosity galaxies \citep{2015ApJ...809L..21M}.} For the lower limit of the size of the stellar body, we used the lower limit on the size of the \HI{} extent. In Fig.~\ref{fig:artificial}, we present a grid of such artificial galaxies. It is evident that in starting from an effective surface brightness of 29.1\,mag\,arcsec$^{-2}$ in the $V$-band, the artificial dwarf galaxies become indistinguishable from the background. This is similar to the surface brightness limit we estimated {previously}.}

Using Fig.~\ref{fig:artificial}, we estimated that a reasonable upper limit on the apparent $V$-band magnitude of the dark cloud is 19.0\,mag. We neglected the small uncertainty of the photometric calibration, {as well as} the foreground extinction of 0.06\,mag given by the NASA/IPAC Extragalactic Database\footnote{\url{https://ned.ipac.caltech.edu/}}.  At the adopted distance, the $V$-band luminosity of the dark cloud {is estimated to} be below $1.1\times10^{7}\,L_\sun$.

{In order to estimate the upper limit on the stellar mass of \obj, we have to multiply the upper limit on the luminosity by the upper limit on the stellar mass-to-light ratio ($M/L$). We derived the latter from the work of \citet{herrmann16}. These authors published $M/L$s for 34 well-studied gas-rich dwarf irregular galaxies from the LITTLE THINGS sample. When estimating the mass of \obj, we used the highest $V$-band $M/L$ that they encountered in their sample, that is to say 1.3. This led us to the upper limit on the stellar mass of \obj of $1.4\times 10^7\,M_\sun$ and to the lower limit on the \HI{}-to-stellar mass ratio of 3.1. As a consistency check, we used the catalog of pressure supported, gas-poor objects from \citet{2016MNRAS.460.4492D} for objects with $L<1.1\times10^7\,L_\sun$. Out of these objects, 90\%  have a $M/L$ below 3, giving a lower estimate of the \HI{}-to-stellar mass ratio for \obj of 1.3.}

\begin{figure*}
\centering
\includegraphics[width=0.99\linewidth]{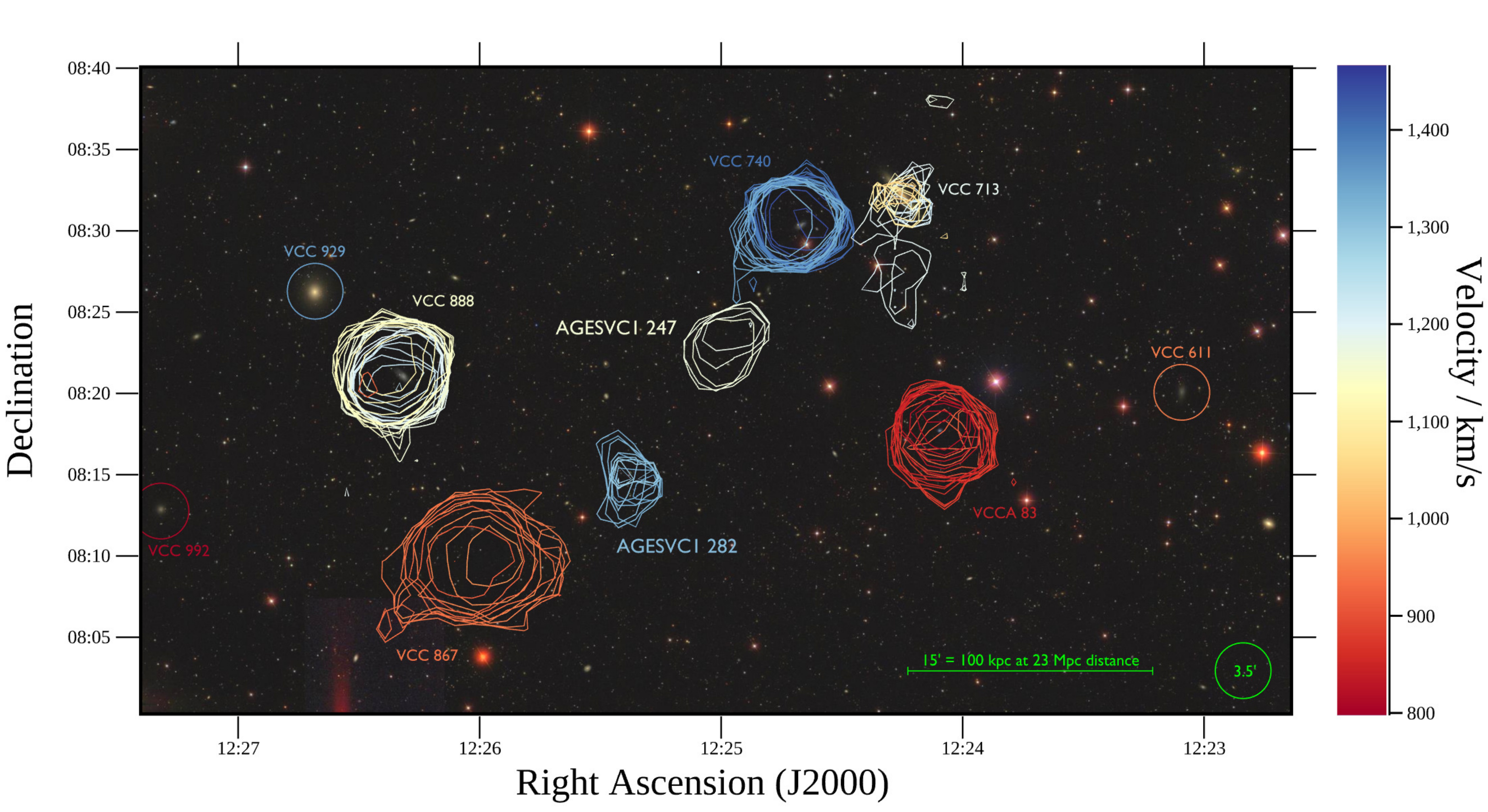}
\caption{Renzogram of the region surrounding AGESVC1 282, overlaid on an optical image from the SDSS. Each contour is at 4$\sigma$ {(where the typical $rms$ is 0.6-0.7 mJy)}, with the color indicating the velocity {(channel width is 5.5\,\kms{})}. Galaxies undetected in \HI{} are marked with circles of the same size as the Arecibo beam.}
\label{fig:cloudenvironment}
\end{figure*}

\section{Discussion and conclusions}
\label{sect:disc}
The dark cloud \obj is one of several \HI{} sources associated with the Virgo cluster without optical counterparts in public surveys. The nature of these sources is {still} unclear. In order to {further constrain} the nature of these objects, we {undertook very deep imaging} of \obj. By not detecting its optical counterpart, we {found that the optical luminosity of the object is below $1.1\times10^{7}\,L_\sun$.} This can be helpful as a constraint for the theoretical models of the formation of the dark clouds or for planning future observations. With the aim to help with the development of the theories of origin of the dark clouds, we briefly discuss several of them for the specific case of \obj on the basis of the \HI{} observations by AGES.

{\bf Tidal debris.} Figure \ref{fig:cloudenvironment} shows the \HI{} contours in the region surrounding \obj, which are overlaid on an SDSS optical image. {Several candidate parent galaxies are visible, though (arguably) none are particularly convincing. The two most promising are VCC 740 and VCC 713 (see \citealt{agestails} for full details). VCC 740 shows a hint of an extension toward \obj{} and it is at a similar radial velocity (66\,\kms{} difference), though the extension is only barely visible in one channel. VCC 713 has a much clearer extension and the mass of \HI{} in the extension is comparable to that in \obj{}. The velocity difference of 195\,\kms{} is compatible with tidal encounters, and while the projected separation of 153\,kpc would make this a large tidal system, it would not be exceptional.}

{While tidal debris is a promising explanation, several difficulties remain (see \citealp{t16} for an extensive discussion). As discussed, the lack of a connecting stream between \obj{} and its parent galaxy is observationally unusual given its high line width. Moreover, the simulations of \citet{t16} and \citet{t17}, looking for clouds matching these parameters, formed through multibody tidal encounters, while omitting the intercluster medium disfavors this scenario.\ That is to say clouds which are both isolated and of a high line width are rare according to the models, and those few that were formed in the usual manner had parent galaxies with much longer detectable \HI{} streams than any seen in this region. We also note that \obj{} is one of six similar features that has been detected by AGES. Finally, the simulated clouds did not deviate as far from the from the BTFR  as the observed features.}

{\bf Dark or low-luminosity galaxy.} The cold dark matter models of galaxy formation {predicts the existence of ``dark galaxies'' with a negligible stellar content}. \cite{t16} simulated the evolution of a dark galaxy corresponding to the observed clouds within the Virgo cluster. They showed that such an object would be {strongly resistant} to the effect of galaxy harassment on {timescales} of a few gigayears and, unlike clouds produced by tidal encounters, could maintain {their} deviation from the BTFR for similar periods. On the other hand, the simulations did not include the inter-cluster medium (ICM) whose effects, such as ram-pressure stripping, might decrease the observability period of the dark galaxies dramatically.  In addition, observations suggest that low-luminosity objects tend to deviate from the BTFR in the opposite direction \citep{2010ApJ...722..248M,2020ApJ...895L...4D,2020arXiv200604606M,2020MNRAS.495.3636M}. It is also worth noting that {in} this scenario, we would expect similar objects to {also exist in other clusters, but the current lack of such detections} might be an observational bias.

{\bf Ram-pressure stripped gas.} Gas can also become detached from galaxies in clusters by ram-pressure stripping. In such a scenario, we would expect the \HI{} signal to form a long band with the source galaxy at one end (see, e.g., \citealp{2014ApJ...792...11J,2014ApJ...780..119K}). Figure \ref{fig:cloudenvironment} suggests that \obj, the dark cloud AGESVC1~247, and the galaxy VCC\,740 indeed lie along a line. We note two difficulties with this explanation. First, the band is not contiguous as in the confirmed cases of ram-pressure stripping. Second, in contradiction with the observed situation, we would expect a monotonic gradient of radial velocity along the gas stream.  

{\bf Magnetically bound gas cloud.} \cite{NonMagneticBlobs} and \cite{MagneticBlobs} describe populations of optically dark neutral gas clouds in cluster simulations, which formed from the cooling and condensation of the hot ICM. {In the latter case,} the pressure from the magnetic fields within these clouds dominates over their thermal pressure and they can survive on gigayear timescales. Unfortunately we cannot compare their kinematics to the observed Virgo clouds since the simulations do not produce artificial spectra. However, while their size (in excess of 1\,kpc) is consistent with the Virgo clouds, the simulated clouds are at least two orders of magnitude less massive -- though the parent halo studied is also less massive than the Virgo cluster. 

{Although there are only} a few known observational characteristics of \obj, we do not have a theoretical explanation of the nature of the object that would fit them all. Further theoretical and observational investigations of the Virgo dark clouds are obviously needed.

This work used the following scientific software: Astrometry.net \citep{2010AJ....139.1782L}, and  SWARP \citep{2010ascl.soft10068B}.\ It also employed the following python3 packages: Astropy \citep{astropy:2013}, ccdproc \citep{2015ascl.soft10007C}, and sep \citep{2016JOSS....1...58B}, which is a python implementation of  Source Extractor \citep{1996A&AS..117..393B}.

\begin{acknowledgements}
{We thank the referee for the constructive report, which helped to clarify and improve the manuscript.}
O.M. is grateful to the Swiss National Science Foundation for financial support. A.V. acknowledges the financial support of the Ministry of Education, Science and Technological Development of the Republic of Serbia (MESTDRS) through the contract No 451-03-68/2020-14/200002 and the financial support by the European Commission through project BELISSIMA (BELgrade  Initiative  for Space  Science,  Instrumentation  and  Modelling  in Astrophysics,  call  FP7-REGPOT-2010-5,  contract  No.  256772), which was used to procure the Milankovi\'c 1.4 meter telescope with the support from the MESTDRS.
R.T. acknowledges support from the Czech Ministry of Education, Youth and Sports from the large Infrastructures
for Research, Experimental Development and Innovations project LM 2015067, the Czech Science Foundation grant CSF 19-18647S, and the institutional project RVO 67985815.
We thank the technical operators at the Astronomical Station  Vidojevica (ASV), Miodrag Sekuli\'c and Petar Kosti\'c, for their excellent work.
\end{acknowledgements}

\bibliographystyle{aa}
\bibliography{literature}

\begin{thebibliography}{60}
\expandafter\ifx\csname natexlab\endcsname\relax\def\natexlab#1{#1}\fi

\bibitem[{{Alam} {et~al.}(2015){Alam}, {Albareti}, {Allende Prieto}, {Anders},
  {Anderson}, {Anderton}, {Andrews}, {Armengaud}, {Aubourg}, {Bailey}, {Basu},
  {Bautista}, {Beaton}, {Beers}, {Bender}, {Berlind}, {Beutler}, {Bhardwaj},
  {Bird}, {Bizyaev}, {Blake}, {Blanton}, {Blomqvist}, {Bochanski}, {Bolton},
  {Bovy}, {Shelden Bradley}, {Brandt}, {Brauer}, {Brinkmann}, {Brown},
  {Brownstein}, {Burden}, {Burtin}, {Busca}, {Cai}, {Capozzi}, {Carnero
  Rosell}, {Carr}, {Carrera}, {Chambers}, {Chaplin}, {Chen}, {Chiappini},
  {Chojnowski}, {Chuang}, {Clerc}, {Comparat}, {Covey}, {Croft}, {Cuesta},
  {Cunha}, {da Costa}, {Da Rio}, {Davenport}, {Dawson}, {De Lee}, {Delubac},
  {Deshpande}, {Dhital}, {Dutra-Ferreira}, {Dwelly}, {Ealet}, {Ebelke},
  {Edmondson}, {Eisenstein}, {Ellsworth}, {Elsworth}, {Epstein}, {Eracleous},
  {Escoffier}, {Esposito}, {Evans}, {Fan}, {Fern{\'a}ndez-Alvar}, {Feuillet},
  {Filiz Ak}, {Finley}, {Finoguenov}, {Flaherty}, {Fleming}, {Font-Ribera},
  {Foster}, {Frinchaboy}, {Galbraith-Frew}, {Garc{\'\i}a},
  {Garc{\'\i}a-Hern{\'a}ndez}, {Garc{\'\i}a P{\'e}rez}, {Gaulme}, {Ge},
  {G{\'e}nova-Santos}, {Georgakakis}, {Ghezzi}, {Gillespie}, {Girardi},
  {Goddard}, {Gontcho}, {Gonz{\'a}lez Hern{\'a}ndez}, {Grebel}, {Green},
  {Grieb}, {Grieves}, {Gunn}, {Guo}, {Harding}, {Hasselquist}, {Hawley},
  {Hayden}, {Hearty}, {Hekker}, {Ho}, {Hogg}, {Holley-Bockelmann}, {Holtzman},
  {Honscheid}, {Huber}, {Huehnerhoff}, {Ivans}, {Jiang}, {Johnson},
  {Kinemuchi}, {Kirkby}, {Kitaura}, {Klaene}, {Knapp}, {Kneib}, {Koenig},
  {Lam}, {Lan}, {Lang}, {Laurent}, {Le Goff}, {Leauthaud}, {Lee}, {Lee},
  {Licquia}, {Liu}, {Long}, {L{\'o}pez-Corredoira}, {Lorenzo-Oliveira},
  {Lucatello}, {Lundgren}, {Lupton}, {Mack}, {Mahadevan}, {Maia}, {Majewski},
  {Malanushenko}, {Malanushenko}, {Manchado}, {Manera}, {Mao}, {Maraston},
  {Marchwinski}, {Margala}, {Martell}, {Martig}, {Masters}, {Mathur},
  {McBride}, {McGehee}, {McGreer}, {McMahon}, {M{\'e}nard}, {Menzel},
  {Merloni}, {M{\'e}sz{\'a}ros}, {Miller}, {Miralda-Escud{\'e}}, {Miyatake},
  {Montero-Dorta}, {More}, {Morganson}, {Morice-Atkinson}, {Morrison},
  {Mosser}, {Muna}, {Myers}, {Nand ra}, {Newman}, {Neyrinck}, {Nguyen},
  {Nichol}, {Nidever}, {Noterdaeme}, {Nuza}, {O'Connell}, {O'Connell},
  {O'Connell}, {Ogando}, {Olmstead}, {Oravetz}, {Oravetz}, {Osumi}, {Owen},
  {Padgett}, {Padmanabhan}, {Paegert}, {Palanque-Delabrouille}, {Pan},
  {Parejko}, {P{\^a}ris}, {Park}, {Pattarakijwanich}, {Pellejero-Ibanez},
  {Pepper}, {Percival}, {P{\'e}rez-Fournon}, {Ṕrez-Ra`fols}, {Petitjean},
  {Pieri}, {Pinsonneault}, {Porto de Mello}, {Prada}, {Prakash},
  {Price-Whelan}, {Protopapas}, {Raddick}, {Rahman}, {Reid}, {Rich}, {Rix},
  {Robin}, {Rockosi}, {Rodrigues}, {Rodr{\'\i}guez-Torres}, {Roe}, {Ross},
  {Ross}, {Rossi}, {Ruan}, {Rubi{\~n}o-Mart{\'\i}n}, {Rykoff},
  {Salazar-Albornoz}, {Salvato}, {Samushia}, {S{\'a}nchez}, {Santiago},
  {Sayres}, {Schiavon}, {Schlegel}, {Schmidt}, {Schneider}, {Schultheis},
  {Schwope}, {Sc{\'o}ccola}, {Scott}, {Sellgren}, {Seo}, {Serenelli}, {Shane},
  {Shen}, {Shetrone}, {Shu}, {Silva Aguirre}, {Sivarani}, {Skrutskie},
  {Slosar}, {Smith}, {Sobreira}, {Souto}, {Stassun}, {Steinmetz}, {Stello},
  {Strauss}, {Streblyanska}, {Suzuki}, {Swanson}, {Tan}, {Tayar}, {Terrien},
  {Thakar}, {Thomas}, {Thomas}, {Thompson}, {Tinker}, {Tojeiro}, {Troup},
  {Vargas-Maga{\~n}a}, {Vazquez}, {Verde}, {Viel}, {Vogt}, {Wake}, {Wang},
  {Weaver}, {Weinberg}, {Weiner}, {White}, {Wilson}, {Wisniewski},
  {Wood-Vasey}, {Ye`che}, {York}, {Zakamska}, {Zamora}, {Zasowski}, {Zehavi},
  {Zhao}, {Zheng}, {Zhou}, {Zhou}, {Zou}, \& {Zhu}}]{2015ApJS..219...12A}
{Alam}, S., {Albareti}, F.~D., {Allende Prieto}, C., {et~al.} 2015, \apjs, 219,
  12

\bibitem[{{Arrigoni Battaia} {et~al.}(2012){Arrigoni Battaia}, {Gavazzi},
  {Fumagalli}, {Boselli}, {Boissier}, {Cortese}, {Heinis}, {Ferrarese},
  {C{\^o}t{\'e}}, {Mihos}, {Cuilland re}, {Duc}, {Durrell}, {Gwyn},
  {Jord{\'a}n}, {Liu}, {Peng}, \& {Mei}}]{vcc1249}
{Arrigoni Battaia}, F., {Gavazzi}, G., {Fumagalli}, M., {et~al.} 2012, \aap,
  543, A112

\bibitem[{{Astropy Collaboration} {et~al.}(2013){Astropy Collaboration},
  {Robitaille}, {Tollerud}, {Greenfield}, {Droettboom}, {Bray}, {Aldcroft},
  {Davis}, {Ginsburg}, {Price-Whelan}, {Kerzendorf}, {Conley}, {Crighton},
  {Barbary}, {Muna}, {Ferguson}, {Grollier}, {Parikh}, {Nair}, {Unther},
  {Deil}, {Woillez}, {Conseil}, {Kramer}, {Turner}, {Singer}, {Fox}, {Weaver},
  {Zabalza}, {Edwards}, {Azalee Bostroem}, {Burke}, {Casey}, {Crawford},
  {Dencheva}, {Ely}, {Jenness}, {Labrie}, {Lim}, {Pierfederici}, {Pontzen},
  {Ptak}, {Refsdal}, {Servillat}, \& {Streicher}}]{astropy:2013}
{Astropy Collaboration}, {Robitaille}, T.~P., {Tollerud}, E.~J., {et~al.} 2013,
  aap, 558, A33

\bibitem[{{Auld} {et~al.}(2006){Auld}, {Minchin}, {Davies}, {Catinella}, {van
  Driel}, {Henning}, {Linder}, {Momjian}, {Muller}, {O'Neil}, {Sabatini},
  {Schneider}, {Bothun}, {Cortese}, {Disney}, {Hoffman}, {Putman}, {Rosenberg},
  {Baes}, {de Blok}, {Boselli}, {Brinks}, {Brosch}, {Irwin}, {Karachentsev},
  {Kilborn}, {Koribalski}, \& {Spekkens}}]{auld}
{Auld}, R., {Minchin}, R.~F., {Davies}, J.~I., {et~al.} 2006, \mnras, 371, 1617

\bibitem[{{Bacchini} {et~al.}(2019){Bacchini}, {Fraternali}, {Iorio}, \&
  {Pezzulli}}]{bacch}
{Bacchini}, C., {Fraternali}, F., {Iorio}, G., \& {Pezzulli}, G. 2019, \aap,
  622, A64

\bibitem[{{Barbary}(2016)}]{2016JOSS....1...58B}
{Barbary}, K. 2016, The Journal of Open Source Software, 1, 58

\bibitem[{{Bertin}(2010)}]{2010ascl.soft10068B}
{Bertin}, E. 2010, {SWarp: Resampling and Co-adding FITS Images Together},
  Astrophysics Source Code Library

\bibitem[{{Bertin} \& {Arnouts}(1996)}]{1996A&AS..117..393B}
{Bertin}, E. \& {Arnouts}, S. 1996, \aaps, 117, 393

\bibitem[{{Bothun} {et~al.}(1987){Bothun}, {Impey}, {Malin}, \&
  {Mould}}]{malin1}
{Bothun}, G.~D., {Impey}, C.~D., {Malin}, D.~F., \& {Mould}, J.~R. 1987, \aj,
  94, 23

\bibitem[{{Collins} {et~al.}(2020){Collins}, {Tollerud}, {Rich}, {Ibata},
  {Martin}, {Chapman}, {Gilbert}, \& {Preston}}]{2020MNRAS.491.3496C}
{Collins}, M. L.~M., {Tollerud}, E.~J., {Rich}, R.~M., {et~al.} 2020, \mnras,
  491, 3496

\bibitem[{{Craig} {et~al.}(2015){Craig}, {Crawford}, {Deil}, {Gomez},
  {G{\"u}nther}, {Heidt}, {Horton}, {Karr}, {Nelson}, {Ninan}, {Pattnaik},
  {Rol}, {Schoenell}, {Seifert}, {Singh}, {Sipocz}, {Stotts}, {Streicher},
  {Tollerud}, {Walker}, \& {ccdproc contributors}}]{2015ascl.soft10007C}
{Craig}, M.~W., {Crawford}, S.~M., {Deil}, C., {et~al.} 2015, {ccdproc: CCD
  data reduction software}

\bibitem[{{Dabringhausen} \& {Fellhauer}(2016)}]{2016MNRAS.460.4492D}
{Dabringhausen}, J. \& {Fellhauer}, M. 2016, \mnras, 460, 4492

\bibitem[{{Danieli} {et~al.}(2020){Danieli}, {van Dokkum}, {Abraham}, {Conroy},
  {Dolphin}, \& {Romanowsky}}]{2020ApJ...895L...4D}
{Danieli}, S., {van Dokkum}, P., {Abraham}, R., {et~al.} 2020, \apjl, 895, L4

\bibitem[{{Davies} {et~al.}(2006){Davies}, {Disney}, {Minchin}, {Auld}, \&
  {Smith}}]{dgmodels}
{Davies}, J.~I., {Disney}, M.~J., {Minchin}, R.~F., {Auld}, R., \& {Smith}, R.
  2006, \mnras, 368, 1479

\bibitem[{{Duc} \& {Bournaud}(2008)}]{duc}
{Duc}, P.-A. \& {Bournaud}, F. 2008, \apj, 673, 787

\bibitem[{{Gavazzi} {et~al.}(1999){Gavazzi}, {Boselli}, {Scodeggio}, {Pierini},
  \& {Belsole}}]{gv99}
{Gavazzi}, G., {Boselli}, A., {Scodeggio}, M., {Pierini}, D., \& {Belsole}, E.
  1999, \mnras, 304, 595

\bibitem[{{Greco} {et~al.}(2018){Greco}, {Greene}, {Strauss}, {Macarthur},
  {Flowers}, {Goulding}, {Huang}, {Kim}, {Komiyama}, {Leauthaud}, {Leisman},
  {Lupton}, {Sif{\'o}n}, \& {Wang}}]{2018ApJ...857..104G}
{Greco}, J.~P., {Greene}, J.~E., {Strauss}, M.~A., {et~al.} 2018, \apj, 857,
  104

\bibitem[{{Green}(2011)}]{2011BASI...39..289G}
{Green}, D.~A. 2011, Bulletin of the Astronomical Society of India, 39, 289

\bibitem[{{Habas} {et~al.}(2020){Habas}, {Marleau}, {Duc}, {Durrell}, {Paudel},
  {Poulain}, {S{\'a}nchez-Janssen}, {Sreejith}, {Ramasawmy}, {Stemock},
  {Leach}, {Cuillandre}, {Gwyn}, {Agnello}, {B{\'\i}lek}, {Fensch},
  {M{\"u}ller}, {Peng}, \& {van der Burg}}]{2020MNRAS.491.1901H}
{Habas}, R., {Marleau}, F.~R., {Duc}, P.-A., {et~al.} 2020, \mnras, 491, 1901

\bibitem[{{Herrmann} {et~al.}(2016){Herrmann}, {Hunter}, {Zhang}, \&
  {Elmegreen}}]{herrmann16}
{Herrmann}, K.~A., {Hunter}, D.~A., {Zhang}, H.-X., \& {Elmegreen}, B.~G. 2016,
  \aj, 152, 177

\bibitem[{{Hess} {et~al.}(2017){Hess}, {Cluver}, {Yahya}, {Leisman}, {Serra},
  {Lucero}, {Passmoor}, \& {Carignan}}]{hess}
{Hess}, K.~M., {Cluver}, M.~E., {Yahya}, S., {et~al.} 2017, \mnras, 464, 957

\bibitem[{{Impey} {et~al.}(1990){Impey}, {Bothun}, {Malin}, \&
  {Staveley-Smith}}]{imp}
{Impey}, C., {Bothun}, G., {Malin}, D., \& {Staveley-Smith}, L. 1990, \apjl,
  351, L33

\bibitem[{{J{\'a}chym} {et~al.}(2014){J{\'a}chym}, {Combes}, {Cortese}, {Sun},
  \& {Kenney}}]{2014ApJ...792...11J}
{J{\'a}chym}, P., {Combes}, F., {Cortese}, L., {Sun}, M., \& {Kenney}, J. D.~P.
  2014, \apj, 792, 11

\bibitem[{{Javanmardi} {et~al.}(2016){Javanmardi}, {Martinez-Delgado},
  {Kroupa}, {Henkel}, {Crawford}, {Teuwen}, {Gabany}, {Hanson}, {Chonis}, \&
  {Neyer}}]{2016A&A...588A..89J}
{Javanmardi}, B., {Martinez-Delgado}, D., {Kroupa}, P., {et~al.} 2016, \aap,
  588, A89

\bibitem[{{Jester} {et~al.}(2005){Jester}, {Schneider}, {Richards}, {Green},
  {Schmidt}, {Hall}, {Strauss}, {Vand en Berk}, {Stoughton}, {Gunn},
  {Brinkmann}, {Kent}, {Smith}, {Tucker}, \& {Yanny}}]{2005AJ....130..873J}
{Jester}, S., {Schneider}, D.~P., {Richards}, G.~T., {et~al.} 2005, \aj, 130,
  873

\bibitem[{{Kenney} {et~al.}(2014){Kenney}, {Geha}, {J{\'a}chym}, {Crowl},
  {Dague}, {Chung}, {van Gorkom}, \& {Vollmer}}]{2014ApJ...780..119K}
{Kenney}, J. D.~P., {Geha}, M., {J{\'a}chym}, P., {et~al.} 2014, \apj, 780, 119

\bibitem[{{Kent} {et~al.}(2009){Kent}, {Spekkens}, {Giovanelli}, {Haynes},
  {Momjian}, {Cort{\'e}s}, {Hardy}, \& {West}}]{kent}
{Kent}, B.~R., {Spekkens}, K., {Giovanelli}, R., {et~al.} 2009, \apj, 691, 1595

\bibitem[{{Koda} {et~al.}(2015){Koda}, {Yagi}, {Yamanoi}, \& {Komiyama}}]{koda}
{Koda}, J., {Yagi}, M., {Yamanoi}, H., \& {Komiyama}, Y. 2015, \apjl, 807, L2

\bibitem[{{Koopmann} {et~al.}(2008){Koopmann}, {Giovanelli}, {Haynes}, {Kent},
  {Balonek}, {Brosch}, {Higdon}, {Salzer}, \& {Spector}}]{koop}
{Koopmann}, R.~A., {Giovanelli}, R., {Haynes}, M.~P., {et~al.} 2008, \apjl,
  682, L85

\bibitem[{{Lang} {et~al.}(2010){Lang}, {Hogg}, {Mierle}, {Blanton}, \&
  {Roweis}}]{2010AJ....139.1782L}
{Lang}, D., {Hogg}, D.~W., {Mierle}, K., {Blanton}, M., \& {Roweis}, S. 2010,
  \aj, 139, 1782

\bibitem[{{Leisman} {et~al.}(2016){Leisman}, {Haynes}, {Giovanelli},
  {J{\'o}zsa}, {Adams}, \& {Hess}}]{leisman}
{Leisman}, L., {Haynes}, M.~P., {Giovanelli}, R., {et~al.} 2016, \mnras, 463,
  1692

\bibitem[{{Mancera Pi{\~n}a} {et~al.}(2020){Mancera Pi{\~n}a}, {Fraternali},
  {Oman}, {Adams}, {Bacchini}, {Marasco}, {Oosterloo}, {Pezzulli}, {Posti},
  {Leisman}, {Cannon}, {di Teodoro}, {Gault}, {Haynes}, {Reiter}, {Rhode},
  {Salzer}, \& {Smith}}]{2020MNRAS.495.3636M}
{Mancera Pi{\~n}a}, P.~E., {Fraternali}, F., {Oman}, K.~A., {et~al.} 2020,
  \mnras, 495, 3636

\bibitem[{{McGaugh}(2005)}]{2005ApJ...632..859M}
{McGaugh}, S.~S. 2005, \apj, 632, 859

\bibitem[{{McGaugh} \& {Wolf}(2010)}]{2010ApJ...722..248M}
{McGaugh}, S.~S. \& {Wolf}, J. 2010, \apj, 722, 248

\bibitem[{{Mihos}(2019)}]{2019arXiv190909456M}
{Mihos}, J.~C. 2019, arXiv e-prints, arXiv:1909.09456

\bibitem[{{Mihos} {et~al.}(2018){Mihos}, {Carr}, {Watkins}, {Oosterloo}, \&
  {Harding}}]{mihos}
{Mihos}, J.~C., {Carr}, C.~T., {Watkins}, A.~E., {Oosterloo}, T., \& {Harding},
  P. 2018, \apjl, 863, L7

\bibitem[{{Mihos} {et~al.}(2015){Mihos}, {Durrell}, {Ferrarese}, {Feldmeier},
  {C{\^o}t{\'e}}, {Peng}, {Harding}, {Liu}, {Gwyn}, \&
  {Cuillandre}}]{2015ApJ...809L..21M}
{Mihos}, J.~C., {Durrell}, P.~R., {Ferrarese}, L., {et~al.} 2015, \apjl, 809,
  L21

\bibitem[{{Minchin} {et~al.}(2007){Minchin}, {Davies}, {Disney}, {Grossi},
  {Sabatini}, {Boyce}, {Garcia}, {Impey}, {Jordan}, {Lang}, {Marble},
  {Roberts}, \& {van Driel}}]{VIRGOHI21}
{Minchin}, R., {Davies}, J., {Disney}, M., {et~al.} 2007, \apj, 670, 1056

\bibitem[{{M{\"u}ller} {et~al.}(2017){M{\"u}ller}, {Jerjen}, \&
  {Binggeli}}]{2017A&A...597A...7M}
{M{\"u}ller}, O., {Jerjen}, H., \& {Binggeli}, B. 2017, \aap, 597, A7

\bibitem[{{M{\"u}ller} {et~al.}(2020){M{\"u}ller}, {Marleau}, {Duc}, {Habas},
  {Fensch}, {Emsellem}, {Poulain}, {Lim}, {Agnello}, {Durrell}, {Paudel},
  {S{\'a}nchez-Janssen}, \& {van der Burg}}]{2020arXiv200604606M}
{M{\"u}ller}, O., {Marleau}, F.~R., {Duc}, P.-A., {et~al.} 2020, arXiv
  e-prints, arXiv:2006.04606

\bibitem[{{M{\"u}ller} {et~al.}(2019){M{\"u}ller}, {Vudragovi{\'c}}, \&
  {B{\'\i}lek}}]{2019A&A...632L..13M}
{M{\"u}ller}, O., {Vudragovi{\'c}}, A., \& {B{\'\i}lek}, M. 2019, \aap, 632,
  L13

\bibitem[{{Nelson} {et~al.}(2020){Nelson}, {Sharma}, {Pillepich}, {Springel},
  {Pakmor}, {Weinberger}, {Vogelsberger}, {Marinacci}, \&
  {Hernquist}}]{MagneticBlobs}
{Nelson}, D., {Sharma}, P., {Pillepich}, A., {et~al.} 2020, arXiv e-prints,
  arXiv:2005.09654

\bibitem[{{Prole} {et~al.}(2019){Prole}, {van der Burg}, {Hilker}, \&
  {Davies}}]{2019MNRAS.488.2143P}
{Prole}, D.~J., {van der Burg}, R.~F.~J., {Hilker}, M., \& {Davies}, J.~I.
  2019, \mnras, 488, 2143

\bibitem[{{Rom{\'a}n} \& {Trujillo}(2017)}]{2017MNRAS.468..703R}
{Rom{\'a}n}, J. \& {Trujillo}, I. 2017, \mnras, 468, 703

\bibitem[{{Sandage} \& {Binggeli}(1984)}]{s84}
{Sandage}, A. \& {Binggeli}, B. 1984, \aj, 89, 919

\bibitem[{{Sawala} {et~al.}(2016){Sawala}, {Frenk}, {Fattahi}, {Navarro},
  {Bower}, {Crain}, {Dalla Vecchia}, {Furlong}, {Helly}, {Jenkins}, {Oman},
  {Schaller}, {Schaye}, {Theuns}, {Trayford}, \& {White}}]{2016MNRAS.457.1931S}
{Sawala}, T., {Frenk}, C.~S., {Fattahi}, A., {et~al.} 2016, \mnras, 457, 1931

\bibitem[{{Serra} {et~al.}(2015){Serra}, {Koribalski}, {Kilborn}, {Allison},
  {Amy}, {Ball}, {Bannister}, {Bell}, {Bock}, {Bolton}, {Bowen}, {Boyle},
  {Broadhurst}, {Brodrick}, {Brothers}, {Bunton}, {Chapman}, {Cheng},
  {Chippendale}, {Chung}, {Cooray}, {Cornwell}, {DeBoer}, {Diamond}, {Forsyth},
  {Gough}, {Gupta}, {Hampson}, {Harvey-Smith}, {Hay}, {Hayman}, {Heywood},
  {Hotan}, {Hoyle}, {Humphreys}, {Indermuehle}, {Jacka}, {Jackson}, {Jackson},
  {Jeganathan}, {Johnston}, {Joseph}, {Kamphuis}, {Leach}, {Lenc}, {Lensson},
  {Mackay}, {Marquarding}, {Marvil}, {McClure-Griffiths}, {McConnell}, {Meyer},
  {Mirtschin}, {Neuhold}, {Ng}, {Norris}, {O'Sullivan}, {Pathikulangara},
  {Pearce}, {Phillips}, {Popping}, {Qiao}, {Reynolds}, {Roberts}, {Sault},
  {Schinckel}, {Shaw}, {Shimwell}, {Staveley-Smith}, {Storey}, {Sweetnam},
  {Troup}, {Tzioumis}, {Voronkov}, {Westmeier}, {Whiting}, {Wilson}, {Wong}, \&
  {Wu}}]{serra}
{Serra}, P., {Koribalski}, B., {Kilborn}, V., {et~al.} 2015, \mnras, 452, 2680

\bibitem[{{Simon} \& {Geha}(2007)}]{2007ApJ...670..313S}
{Simon}, J.~D. \& {Geha}, M. 2007, \apj, 670, 313

\bibitem[{{Simpson} {et~al.}(2018){Simpson}, {Grand}, {G{\'o}mez}, {Marinacci},
  {Pakmor}, {Springel}, {Campbell}, \& {Frenk}}]{2018MNRAS.478..548S}
{Simpson}, C.~M., {Grand}, R.~J.~J., {G{\'o}mez}, F.~A., {et~al.} 2018, \mnras,
  478, 548

\bibitem[{{Slater} {et~al.}(2009){Slater}, {Harding}, \&
  {Mihos}}]{2009PASP..121.1267S}
{Slater}, C.~T., {Harding}, P., \& {Mihos}, J.~C. 2009, \pasp, 121, 1267

\bibitem[{{Taylor} {et~al.}(2012){Taylor}, {Davies}, {Auld}, \&
  {Minchin}}]{t12}
{Taylor}, R., {Davies}, J.~I., {Auld}, R., \& {Minchin}, R.~F. 2012, \mnras,
  423, 787

\bibitem[{{Taylor} {et~al.}(2013){Taylor}, {Davies}, {Auld}, {Minchin}, \&
  {Smith}}]{t13}
{Taylor}, R., {Davies}, J.~I., {Auld}, R., {Minchin}, R.~F., \& {Smith}, R.
  2013, \mnras, 428, 459

\bibitem[{{Taylor} {et~al.}(2016){Taylor}, {Davies}, {J{\'a}chym}, {Keenan},
  {Minchin}, {Palou{\v{s}}}, {Smith}, \& {W{\"u}nsch}}]{t16}
{Taylor}, R., {Davies}, J.~I., {J{\'a}chym}, P., {et~al.} 2016, \mnras, 461,
  3001

\bibitem[{{Taylor} {et~al.}(2017){Taylor}, {Davies}, {J{\'a}chym}, {Keenan},
  {Minchin}, {Palou{\v{s}}}, {Smith}, \& {W{\"u}nsch}}]{t17}
{Taylor}, R., {Davies}, J.~I., {J{\'a}chym}, P., {et~al.} 2017, \mnras, 467,
  3648

\bibitem[{{Taylor} {et~al.}(2020){Taylor}, {K{\"o}ppen}, {J{\'a}chym},
  {Minchin}, {Palou{\v{s}}}, \& {W{\"u}nsch}}]{agestails}
{Taylor}, R., {K{\"o}ppen}, J., {J{\'a}chym}, P., {et~al.} 2020, \aj, 159, 218

\bibitem[{{Taylor} {et~al.}(2018){Taylor}, {W{\"u}nsch}, \&
  {Palou{\v{s}}}}]{t18}
{Taylor}, R., {W{\"u}nsch}, R., \& {Palou{\v{s}}}, J. 2018, \mnras, 479, 377

\bibitem[{{Trujillo} \& {Fliri}(2016)}]{2016ApJ...823..123T}
{Trujillo}, I. \& {Fliri}, J. 2016, \apj, 823, 123

\bibitem[{{van Dokkum} {et~al.}(2015){van Dokkum}, {Abraham}, {Merritt},
  {Zhang}, {Geha}, \& {Conroy}}]{vandok}
{van Dokkum}, P.~G., {Abraham}, R., {Merritt}, A., {et~al.} 2015, \apjl, 798,
  L45

\bibitem[{{Venhola} {et~al.}(2017){Venhola}, {Peletier}, {Laurikainen}, {Salo},
  {Lisker}, {Iodice}, {Capaccioli}, {Verdois Kleijn}, {Valentijn}, {Mieske},
  {Hilker}, {Wittmann}, {van de Ven}, {Grado}, {Spavone}, {Cantiello},
  {Napolitano}, {Paolillo}, \& {Falc{\'o}n-Barroso}}]{2017A&A...608A.142V}
{Venhola}, A., {Peletier}, R., {Laurikainen}, E., {et~al.} 2017, \aap, 608,
  A142

\bibitem[{{Villaescusa-Navarro} {et~al.}(2016){Villaescusa-Navarro},
  {Planelles}, {Borgani}, {Viel}, {Rasia}, {Murante}, {Dolag}, {Steinborn},
  {Biffi}, {Beck}, \& {Ragone-Figueroa}}]{NonMagneticBlobs}
{Villaescusa-Navarro}, F., {Planelles}, S., {Borgani}, S., {et~al.} 2016,
  \mnras, 456, 3553

\end{thebibliography}

\end{document}